\documentclass[preprint,showkeys]{revtex4-1}

\usepackage{graphicx}                         
\usepackage{dcolumn}                          
\usepackage[version=3]{mhchem}                
\usepackage[T1]{fontenc}                      

\usepackage{epsfig}
\usepackage{epstopdf}
\usepackage{amsbsy}
\usepackage{multirow}
\usepackage{bm}
\usepackage{caption}
\usepackage{siunitx}
\usepackage {CJK}
\usepackage{textcomp}
\usepackage{comment}
\usepackage[ruled,vlined]{algorithm2e}
\usepackage{caption}
\usepackage{subcaption}
\usepackage{xcolor}
\usepackage{array}

\newcolumntype{P}[1]{>{\centering\arraybackslash}p{#1}}

\begin{document}

\title{Understanding defect structures in metal additive manufacturing via molecular dynamics}
\author{Gurmeet Singh, Anthony M. Waas and Veera Sundararaghavan}
\email{Corresponding author. 
\newline \mbox{\textit{E-mail address}: veeras@umich.edu (V. Sundararaghavan).}}
\affiliation{Department of Aerospace Engineering, University of Michigan, Ann Arbor, MI, U.S.A. 48109}

\begin{abstract}
Additive manufacturing of a single crystalline metallic column is studied using molecular dynamics simulations. In the model, a melt pool is incrementally added and cooled to a target temperature under isobaric conditions to build a metallic column from bottom up. Common neighbor analysis (CNA) is used to observe the evolution of atomic scale defects during this process. The solidification is seen to proceed along two directions for an added molten layer. The molten layer in contact with the cooler lattice has a fast solidification front that competes with the slower solidification front starting from the top layer. Defect structure formed strongly depends on the speeds of the two competing solidification fronts. Up to a critical layer thickness, defect free single crystals are obtained as the faster solidification front reaches the top of the melt pool before the initiation of the slower front from the top. Slower cooling rates lead to reduction in defects, however, the benefits diminish below a critical rate. Defect content can be significantly reduced by raising the temperature of the powder bed to a critical temperature. This temperature is determined by two competing mechanisms: slower cooling rates at higher temperatures competing against increase in amorphousness as one gets closer to the melting point. Finally, effect of added soft inclusion (SiS$_2$) and a hard inclusion (SiC) on the defect structure is studied. Hard inclusions lead to retained defect structure while soft inclusions reduce defective content compared to a pure metal. 
\end{abstract}


\maketitle

\section{Introduction} 

\begin{figure}[h]
    \centering
    \includegraphics[trim={1.0cm 3.6 0.2cm 0cm},clip=true ,width=0.98\textwidth]{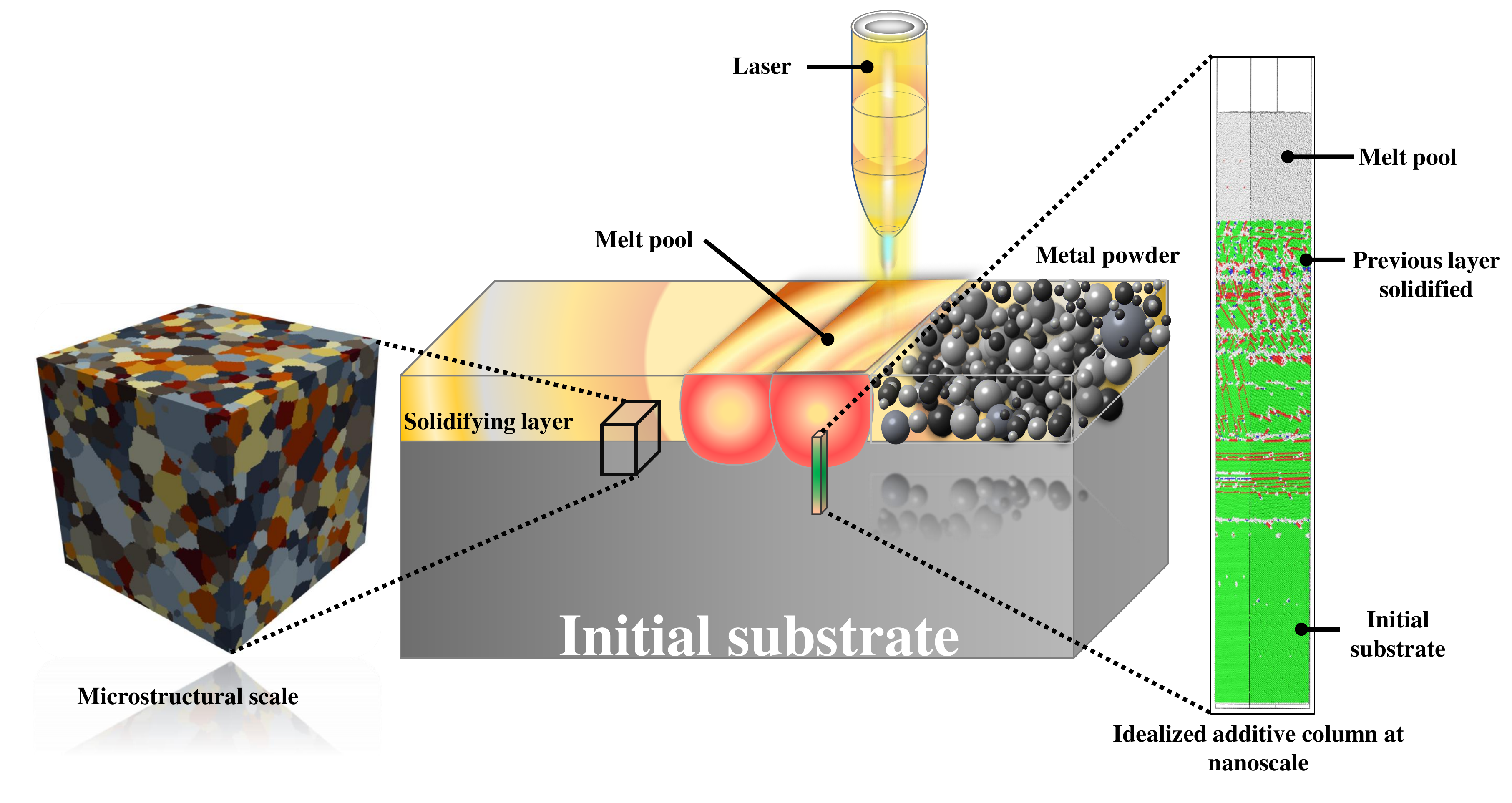}
    \caption{A schematic of the additive manufacturing process (center) depicting the micro-structure scale (left) and the idealized additive process at nanoscale (right).}
    \label{fig:Schematic}
\end{figure}

Additive manufacturing (AM) is now being routinely used in industry to build metal parts due to the flexibility it allows in part design. Additive manufacturing involves gradual increase in the size and shape of solids due to the addition of new material layers on top of existing ones. Thermally enabled metal additive manufacturing (MAM) consists of fast heating  (eg. using lasers) of powder beds leading to large thermal gradients followed by high cooling rates due to the small volumes that are exposed at a time \cite{murr2012metal,Herzog2016}. A resolution of upto $20 ~\mu m$ per layer \cite{Bandyopadhyay2020} and components with functionally graded materials has been realized \cite{Zhang2019}. Defect formation during additive manufacturing is a subject of in--depth studies as it relates to accelerated fatigue and fracture of the respective part \cite{VanSickle2020}. There is a high dependence on several processing parameters which dictates the mechanisms and properties of the final product \cite{Bajaj2020,Herzog2016, Shamsaei2015,Bartlett2019}. Wang et al. studied the influence of the laser scan rate and laser power on the residual stresses in the fabricated samples \cite{Wang2009,wang2008residual}. Their findings revealed that higher laser power led to an increase in the residual stress in the build direction and laser speed influences stresses in transverse directions of the additive samples \cite{wang2008residual}. Fig.\ref{fig:Schematic} shows a schematic of the metal additive manufacturing process (center) with a representative microstructure formed shown. 

In order to understand microstructure evolution and defect formation as a function of process parameters, it is important to understand solidification processes in the melt and how it influences the defect structure in an additive column. Several simulation methods have been applied to study the solidification phenomenon at different scales \cite{Li2020,Thompson2015}. Cellular automata in conjunction with temperature history from a finite element analysis (FEA) \cite{Chen2016}, finite volume \cite{Li2018} or finite difference solutions \cite{Zinovieva2018} have been implemented to predict the grain size and texture during solidification under arc welding or additive manufacturing process. A continuum scale model by Prabhakar et al. \cite{prabhakar2015computational} was implemented that employs FEA and thermal history in order to simulate the deformation and distortion of the part. Another popular choice for simulating structure at meso scales is phase field (PF) modeling \cite{Sahoo2016,Tan2020}. However, because of the length scale and extreme heating and cooling rates, melt solidification is a highly non-equilibrium process and continuum methods alone are not sufficient. 

Molecular dynamics simulations are able to capture the non--equilibrium physics by efficiently representing the inter-atomic interactions during solidification using well calibrated interatomic potentials. Although only small volumes (nanoscales) can be simulated, such simulations have allowed prediction of bulk phenomena in the past including understanding of metal curves \cite{Karavaev2016,Asadi2015}, nucleation and grain growth \cite{Sui2018}, solidification defects \cite{Mahata2019,Shibuta2011} and vacancy formation during solidification \cite{Zhang2017}. Kurian et al. \cite{Kurian2020} recently employed large--scale molecular dynamics simulations to study the melt behavior and its interaction with nano-powdered particles and solidification.  The study showed the process of crystal nucleation in the melt and the emergence of grain boundaries and voids during the process. Another study by Jiang et al. \cite{Jiang2020} employed molecular dynamics simulations to understand the crystallization and cluster evolution patterns for various laser powers and scan rates in Fe$_{50}$Ni$_{50}$ amorphous alloy.  The investigations demonstrated that the low scanning speeds allow increased crystallization into body-centered cubic (BCC) structure. Furthermore, low laser energy density diminishes crystallization and leads to a more amorphous structure \cite{Jiang2020}. However, a study of atomic scale defects as a function of process parameters was not performed in these studies and this is the goal of the current work. 

During solidification of grains, various defect structures form that are either unstable (eg. self--interstitials) which are relieved during reheating or more stable defect structures (eg. high angle grain boundaries) that persist after the process. Understanding the formation of defects as a function of parameters such as cooling rates, particle bed temperature, presence of inclusions etc. has not yet been done via  molecular simulations. Hu et al. experimentally studied the effect of cooling rates on the microstructural characteristics in an MAM process \cite{hu2019towards} and found that fast cooling rates results in small grain size and at slower cooling rates, large grain structures were observed. There have been experimental investigations on the effect of layer thickness by Sui et al. \cite{sui2003investigation}, and on the influence of substrate temperature (preheated bed) \cite{Muller2019} on microstructure evolution and  mechanical performance \cite{Shamsaei2015}. By increasing the substrate temperature to 1000 $^o$C, mitigation of the extent and reduction in the microcrack density was observed in Tungsten as compared to preheated bed at 200 $^o$C. In this study, we analyze the effect of such process parameters by idealizing an additive column at nanoscale using molecular dynamics (MD) as shown in Fig. \ref{fig:Schematic} (right). Fundamental reasons for these effects are analyzed and interpreted at the atomistic scale.

The simulation in this work involves melting and cooling of successive layers of material. The simulation idealizes the process at nano--scales and the metallic column simulated in this work, in relation to the continuum scale additive process, is shown in Fig.\ref{fig:Schematic}(right). We first obtain the equilibrated density of the liquid metal at a temperature above melting point. A liquid melt of this density is then poured onto a solidified substrate in a  layer--by--layer fashion along a build direction (we choose <001> for both Cu and Al systems) followed by cooling down of the system to a target temperature. The study primarily aims at the understanding of the defect structure and its origin as a function of several process variables such as cooling rates, layer thickness and bed temperature from an atomistic viewpoint. Section III presents detailed discussion of interacting phenomena that lead to differentiation of single crystals into dislocation defects or amorphous regions as a function of these process variables. In the latter part of the section, we simulate the effect of inclusions in the melt on the defect structure. Section IV presents the key conclusions drawn from this study.

\begin{figure}[h]
    \centering
    \includegraphics[width=0.8\textwidth]{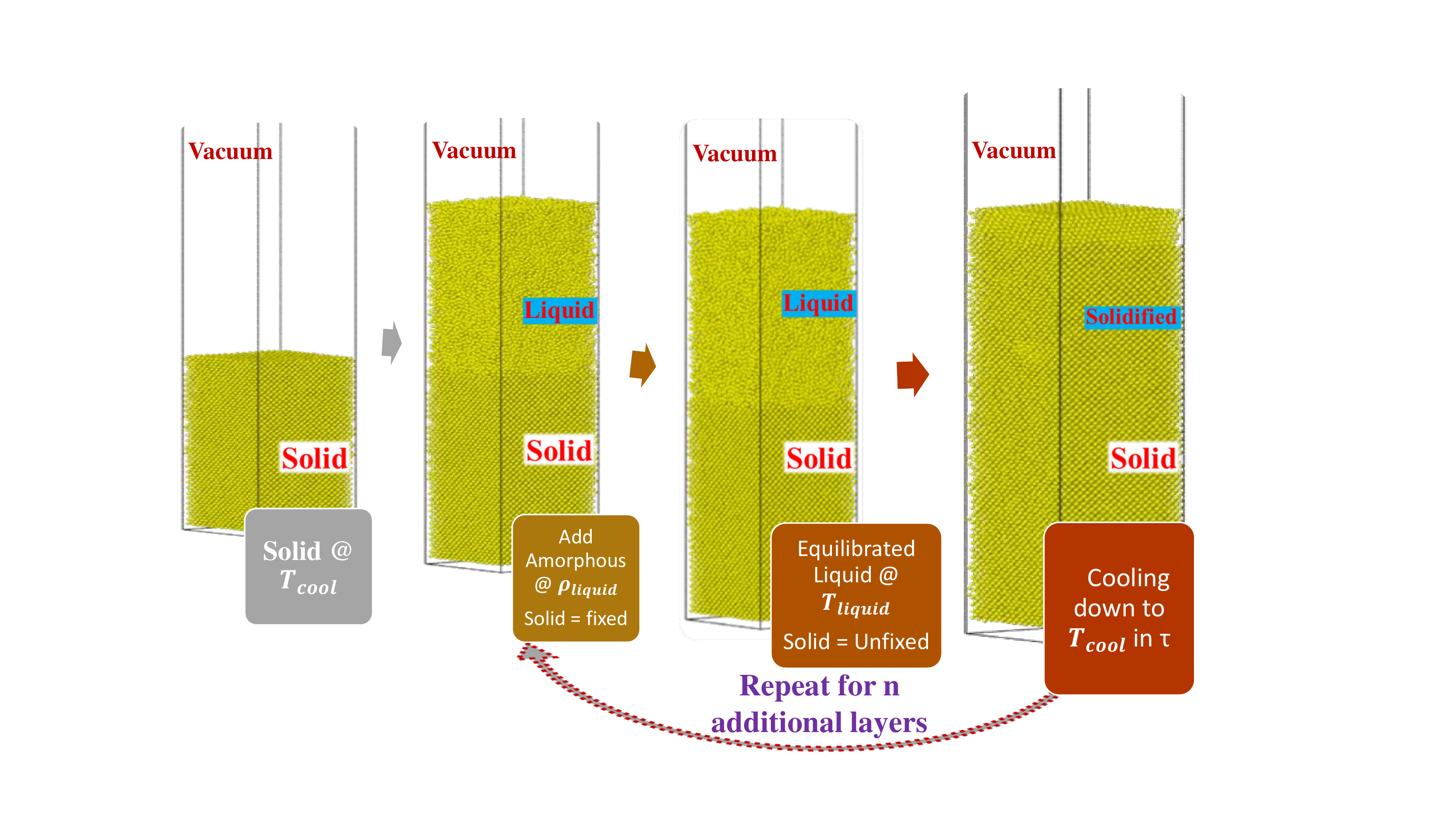}
    \caption{Steps for molecular dynamics simulation of the additive manufacturing process.}
    \label{fig:SimSetup}
\end{figure}

\section{Methods}

All MD simulations are performed using large-scale atomic/molecular massively parallel simulator (LAMMPS) \cite{plimpton1995fast}.  A supercell of copper of user--specified dimensions was generated as the basis for an initial solid seed region with periodicity in all directions. The supercell was initially equilibrated to a target bed temperature ($T_{cool}$) using NPT simulation that employed an Andersen barostat \cite{andersen1980molecular} at atmospheric pressure and a Nosé-Hoover (NH) thermostat \cite{samoletov2007thermostats}. The Andersen barostat only allows isotropic changes to the unit cell and ensures that the FCC lattice has the correct equilibrium volume at room temperature.  In order to estimate the density of the liquid ($\rho_{liquid}$) at the melting temperature, this cell was heated to $T_{liquid}$ using another NPT simulation. The density of the liquid cell was noted and was used while adding liquid layers during the additive process. 

During the deposition simulation, the following process was followed (shown in Fig. \ref{fig:SimSetup}). The equilibrated supercell (at $T_{cool}$) of known thickness ($t$) was initialized at a target temperature ($T_{cool}$) and a vacuum layer was created on top. At each deposition stage, the same volume as the solid supercell was amorphously packed at the top of the solid column at the liquid density $\rho_{liquid}$ estimated previously. The solid atoms were then fixed and an NVT simulation was used to equilibrate the liquid layer to the melt temperature ($T_{liquid}$) using velocity re-scaling. The solid atoms are then unfixed and a cool--down is performed starting from the equilibrated velocities using the NH thermostat and an Andersen barostat at atmospheric pressure. Along the build direction, a vacuum region is maintained to avoid self interactions along the build axis. The entire cell is cooled back to a set temperature ($T_{cool}$) over a specified time ($\tau$). Once cooled back $T_{cool}$, the melt layer equilibrates approximately to a layer thickness ($t$) accounting for contraction during cooling. In the next deposition stage (for the second melt pool), the process is repeated by amorphously packing another liquid layer at the top of the newly solidified structure at the known liquid density $\rho_{liquid}$ and repeating the cooling steps detailed above. 

The layer thickness and cooling times can be controlled in the simulation. Higher substrate temperatures (use of preheated beds) can be simulated by changing the equilibration temperatures set for the thermostat. The following parameters were varied in this work:
\begin{itemize}
    \item Cooling time ($\tau$) of each deposited layer
    \item Thickness of deposited layers ($t$)
    \item Target cooling temperature ($T_{cool}$) of each deposited layer
    \item Influence of a soft (SiS$_2$) and hard (SiC) inclusion in each deposited layer
\end{itemize}

The structure type in the simulation box is identified by performing common neighbor analysis (CNA) \cite{Faken1994} in Ovito \cite{stukowski2009visualization,stukowski2012structure}, an open source visualization tool. Since FCC metals are chosen in this study, we refer to the FCC content in the lattice as the defect--free content for the additive column. In the perfect FCC crystal lattice, three equivalent close-packed planes are aligned along 111 directions which leads to an atomic stacking sequence of the form ABCABCABC. A typical defect is the formation of a stacking fault which occurs when the stacking sequence changes through removal or misalignment of one of the layers in the form ABCABABC.  Such regions are identified to be of hexagonal close packed (HCP) form in the software. Higher energy crystal structures of the form body centered cubic (BCC) are also identified. The clusters that do not fall under the classification of cubic or hexagonal crystals are termed {'}amorphous{'} and may contain defects such as grain boundary dislocations, shockley partials, vacancies and self--interstitials in addition to disordered clusters. The actual nature of these amorphous defects are obtained through observation. These clusters are identified at the end of cooling time for each layer. After $i^{th}$ layer is added to the additive column, the percentage of each structure is computed at layer number $L_i$ using the following equation:

\begin{equation}\label{defect_perc}
    \%FCC^{i} = \frac{N_{FCC}^{i}}{N_{total}^{i}}\times 100 ~~~\mbox{and}~~~\\
    \%Amorphous^{i} = \frac{N_{Amorphous}^{i}}{N_{total}^{i}}\times 100
\end{equation}
where, we refer $i=0$ as the initial solid crystal and for $i=1$ as the first liquid layer  ($L_1$) added and so on. $N_{FCC}^{i}$, $N_{Amorphous}^{i}$ and $N_{total}^{i}$ are number of FCC, amorphous and total number of atoms after the $i^{th}$ layer is added to the additive column, respectively. 

\section{Results and Discussion}

\begin{figure}[h]
     \centering
         \includegraphics[width=0.7\textwidth]{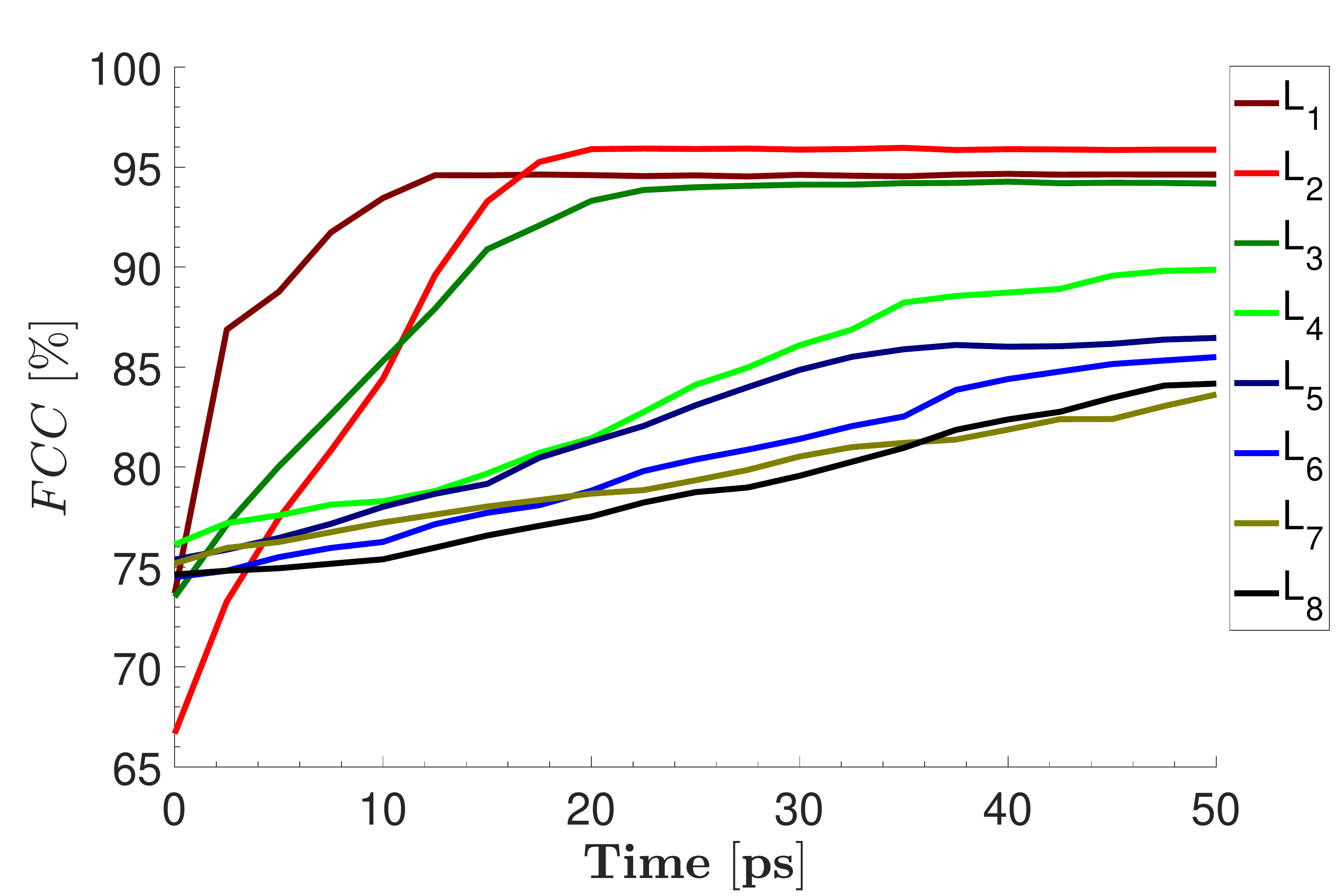}
        \caption{Evolution of FCC structure with cooling time for $16a\times16a\times16a$ system.}
         \label{fig:fcc_Vs_time16x16x16}
\end{figure}

The molecular simulations in these examples were carried out for face centered cubic (FCC) Copper (melting point of 1358 K) using an embedded atom potential (EAM) describing the atomic interactions \cite{zhou2004misfit}.  The lattice constant of copper is $a=3.597$ \r{A} at $T_{cool} = 300$ K. The melt was taken to 1500 K and at this temperature, the density is computed as  $\rho_{liquid}=7.8$~gcm$^{-3}$. Each liquid layer was equilibrated at $T_{liquid}=1500$ K for 25 ps before adding onto the solidified column. The size ($x\times y\times z$) of the initial supercell considered in this case is $16a\times16a\times16a$, with the last number indicating the thickness ($t$) along the build direction ($z$).   Fig. \ref{fig:fcc_Vs_time16x16x16} shows the evolution of FCC content (as given be Eq. \ref{defect_perc}) for each added molten layer during cooling within a time of 50 ps. Eight layers in total were simulated. Initial FCC content is low due to the amorphous nature of the liquid melt but tends to increase as the system solidifies to a crystalline structure at room temperature.

\begin{figure}[h]
    \centering
    \includegraphics[width=\textwidth]{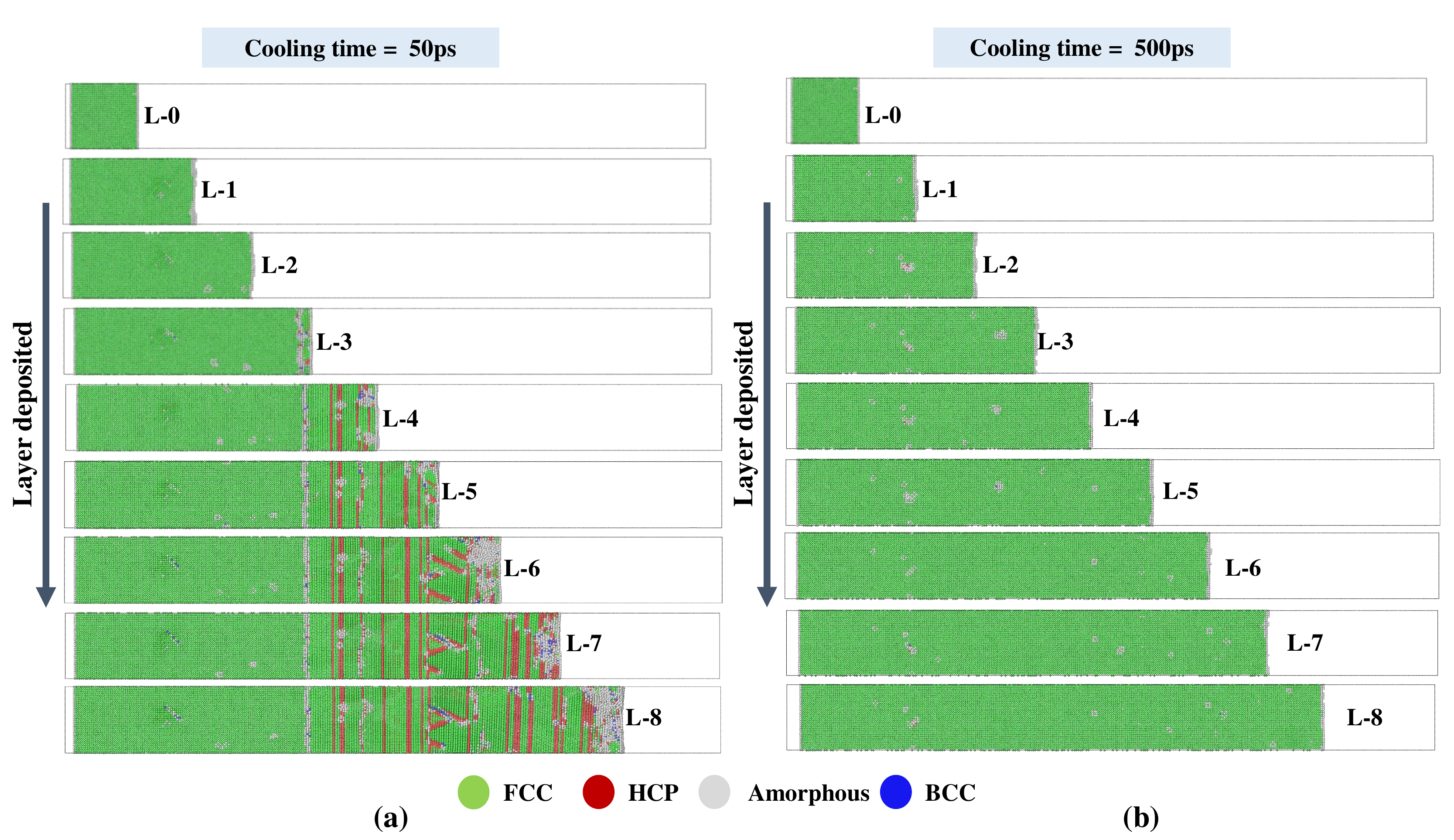}
    \caption{Evolution of defects for $16a$ liquid layer thickness under 50 ps (a) and 500 ps as a function of number of layers deposited.}
    \label{fig:CNA_16x16x16}
\end{figure}

In this figure, it is seen that the first melt layer solidifies rapidly (within 10 ps) from the defect free substrate, but locks in around 5\% defects. Subsequent layers take more and more time to achieve a stable defect profile. This is related to decrease in solidification front speeds as the defects in the substrate increases, which is studied later in this section. As molten layer 2 is added and equilibrated, some of the defects in layer 1 is ameliorated (see animation01 in SI). However, the defect volume fraction continues to increase from layer 3 onwards as sufficient cooling time to eliminate defects is not provided. The defect evolution for this case is shown in Fig. \ref{fig:CNA_16x16x16}(a) showing steady increase in stacking faults (HCP stacking, shown in red) and amorphous regions. However, by providing additional time for defects to rearrange (with cooling time set ten--fold higher at 500 ps), all the stacking faults are eliminated and small amount (2\%) of dislocation defects are formed. 

\begin{figure}[h]
     \centering
     \begin{subfigure}[b]{0.48\textwidth}
         \centering
         \includegraphics[width=\textwidth]{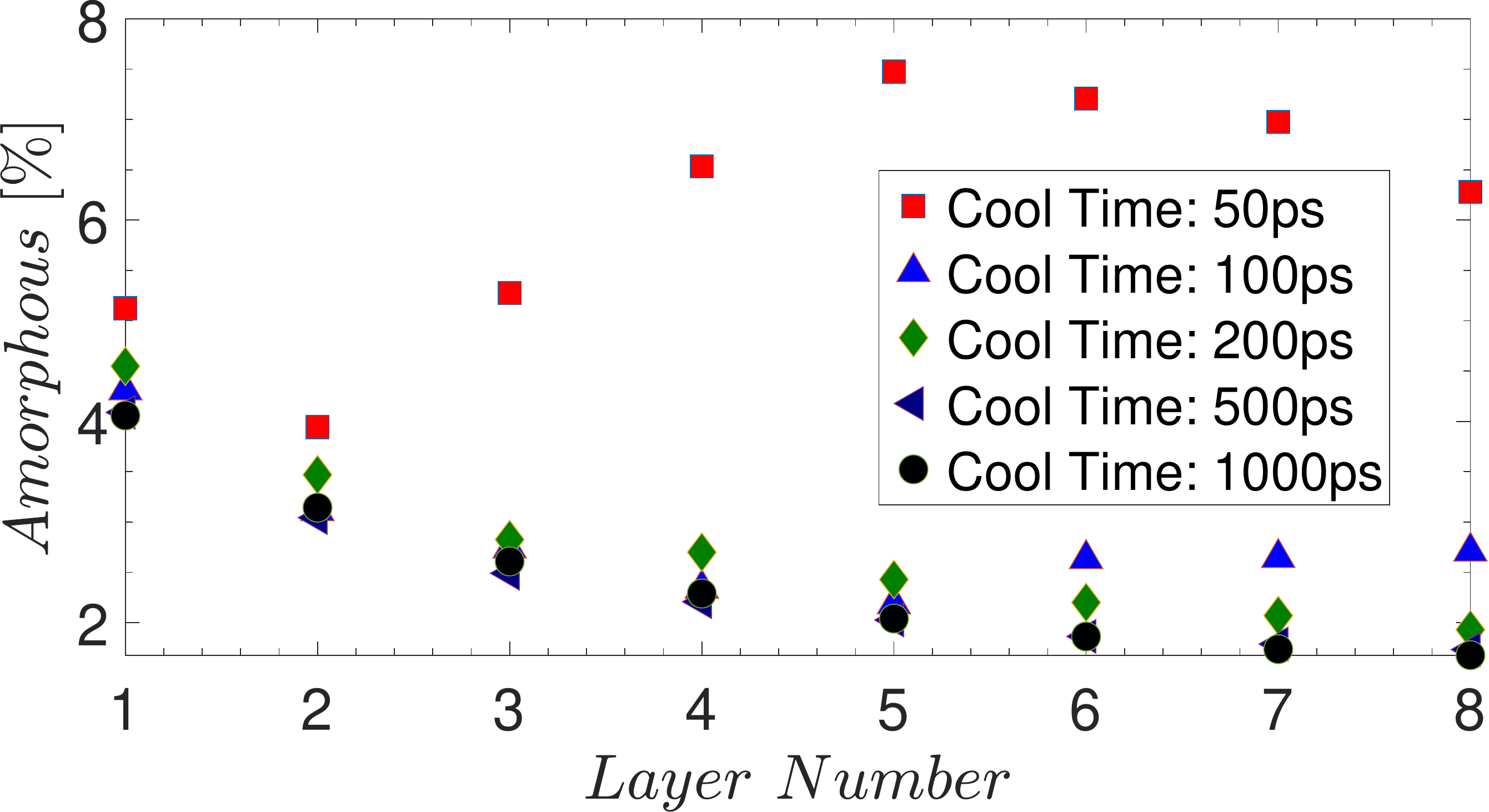}
         \caption{ }
         \label{fig:amorph_Vs_L16x16x16}
     \end{subfigure}
     \hfill
     \begin{subfigure}[b]{0.48\textwidth}
         \centering
         \includegraphics[width=\textwidth]{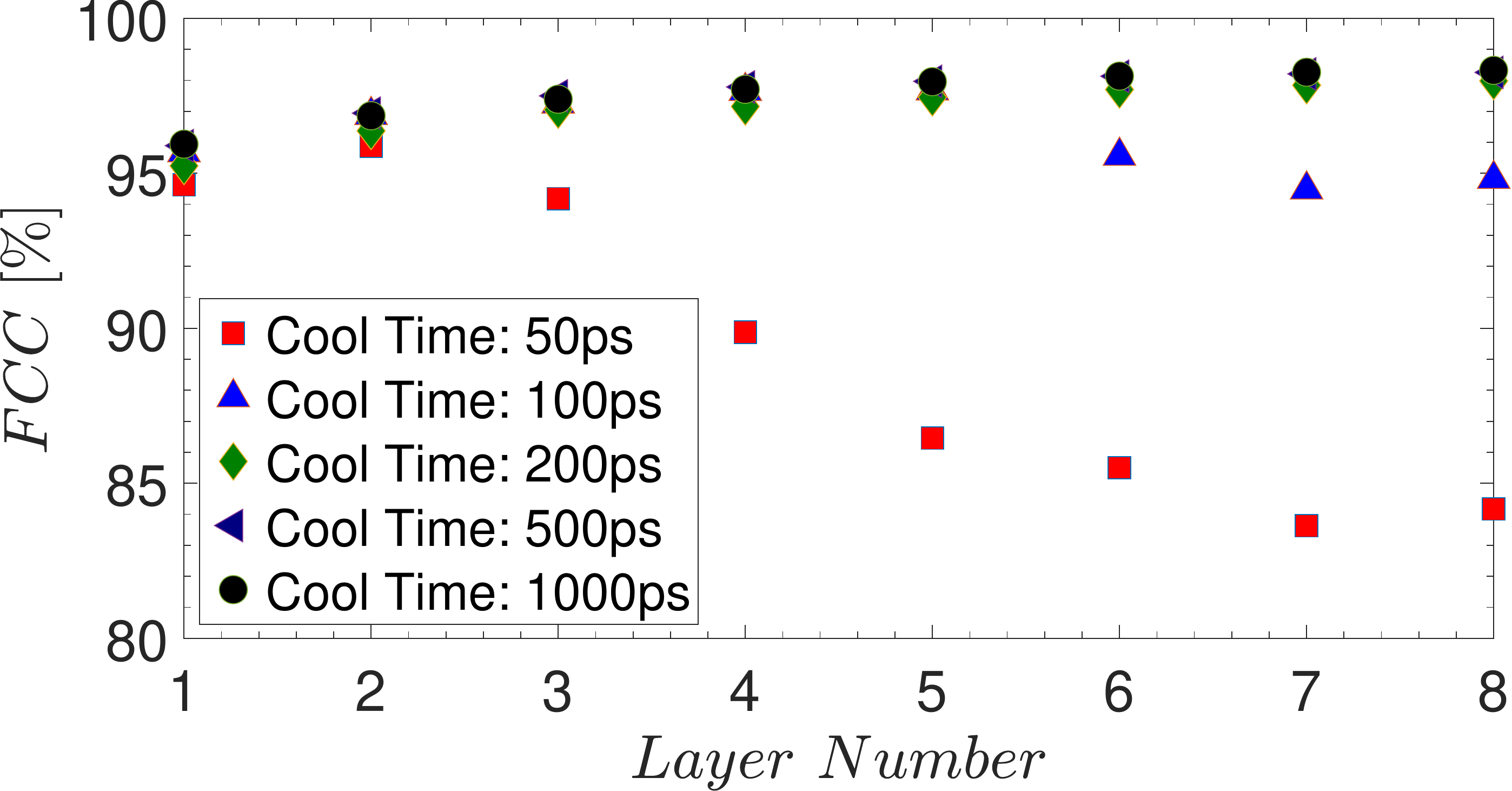}
         \caption{ }
         \label{fig:fcc_Vs_L16x16x16}
     \end{subfigure}
        \caption{Amorphous (a) and FCC (b) structure type evolution as a function of added layers for different cooling times in a $16a\times16a\times16a$ system.}
        \label{fig:struc_Vs_L16x16x16}
\end{figure}

Fig. \ref{fig:struc_Vs_L16x16x16} shows this effect of cooling time on defects in detail. For the 50 ps cooling time detailed previously, the amount of defects increase with the number of layers. However, as the cooling time given to the melt is raised to 100 ps, the defect volume fraction significantly reduces with an increase in number of layers due to sufficient time given for rearrangement and elimination of higher energy defects. However, the benefits of slow cooling is lost after 500 ps as the defects formed are stable and cannot be removed even if the cooling time is increased to 1000 ps (see animation02 for 500 ps case in SI). At 1000 ps cooling time, the percentage of defect--free content converges towards 98\% as the number of layers are increased. 

\subsection{Influence of layer thickness on defects}

In this parametric study, the thickness of melt region is increased keeping the orthogonal dimensions fixed. This simulates larger volumes of molten material added at each step, corresponding to higher energy processing, which affects the solidification kinetics in the build direction ($z-axis$). The number of atoms in the initial solid and for the liquid layer for four of the thicknesses studied here are presented in Table \ref{LayerThickness}. The system is cooled down to $T_{cool}=300$ K at a cooling time ($\tau$) of 500 ps for all cases. The thickness in the build direction ($z$-axis) is varied starting from $10a$ lattice units to  $96a$ keeping the transverse dimensions fixed at $16a\times16a$. The layer thickness, layerwise number of atoms and the total number of atoms (at the end of eighth layer) for some of the thickness cases are shown in Table \ref{LayerThickness}. 

\begin{table}[h]
\centering
\caption{System parameters for different melt layer thickness studies.}\label{LayerThickness} 
\begin{tabular}{P{3.5cm}P{2.0cm}P{3.0cm}P{3.0cm}P{4.5cm}}
\hline
\textbf{System} & \textbf{Thickness ($nm$)} & \textbf{Atoms in initial solid layer ($L_0$)} & \textbf{Atoms in each liquid layer} & \textbf{Maximum number of atoms simulated}\\ \hline
\textbf{$16a\times16a\times16a$} & 5.783 & 16896 & $\sim$14,221 & 132,088 \\ \hline
\textbf{$16a\times16a\times32a$} & 11.565 & 33280 & $\sim$28,284  & 261,663 \\ \hline
\textbf{$16a\times16a\times48a$} & 17.348 & 49664 & $\sim$42,702 & 395,008 \\ \hline
\textbf{$16a\times16a\times96a$} & 32.698 & 49664 & $\sim$85,906 & 777,564 \\ \hline
\end{tabular}
\end{table}

\begin{figure}[h]
    \centering
    \includegraphics[width=\textwidth]{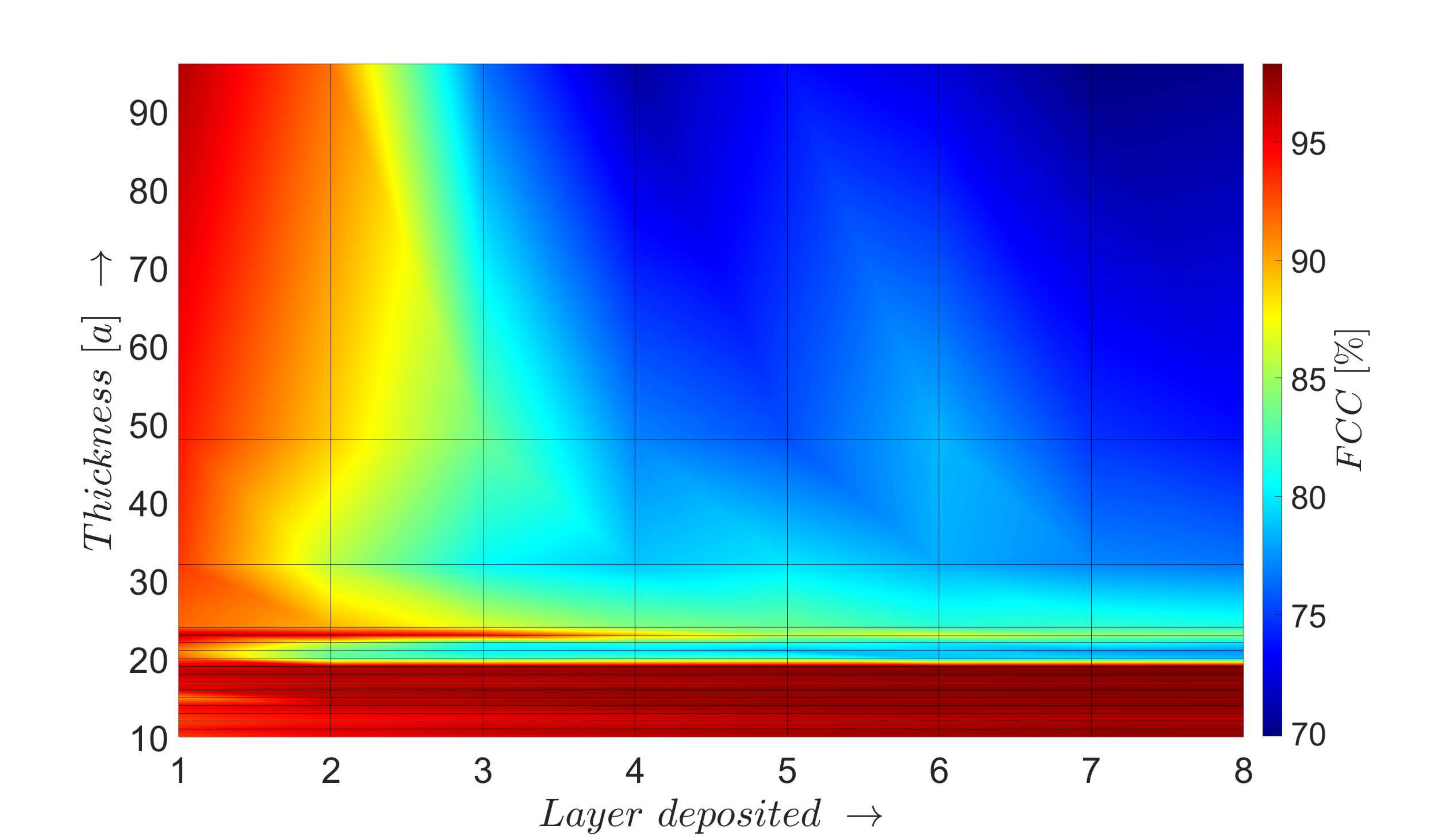}
    \caption{Contour plot of FCC percentage as a function of layer thickness and the number of layers.}
    \label{fig:contour3D_thickness}
\end{figure}

Fig. \ref{fig:contour3D_thickness} shows the FCC content in the cooled down layers (on horizontal axis) for different melt thickness (vertical axis). We can clearly see that the FCC fraction is converged at layer 4 to layer 5 for all the thicknesses studied. For lower melt thicknesses, we see single crystal growth with no significant defect structure. However, at higher thicknesses of added layers, significant amount of defects emerge. A clear threshold thickness beyond which the trend of perfect crystallization changes to a defective structure is seen. For the case of Cu at 500 ps cooling time, this threshold value is $20a$ at and beyond which the defects starts initiating and evolving with additional layers to the additive column.   

\begin{figure}[h]
    \centering
    \includegraphics[width=\textwidth]{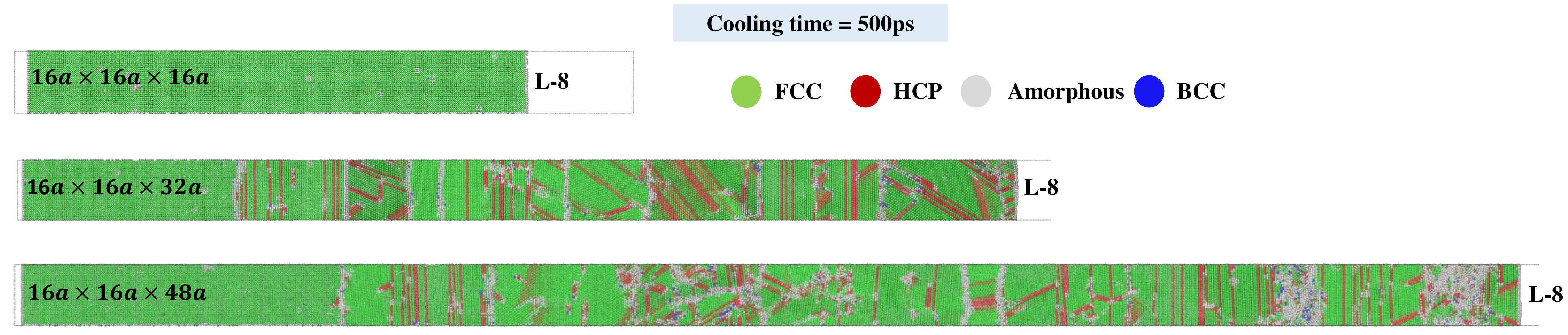}
    \caption{Snapshot of the structures produced for three different deposited layer thicknesses at the end of deposition of eight layers.}
    \label{fig:CNA_thickness}
\end{figure}

Fig. \ref{fig:CNA_thickness} shows defect structure at the end of addition of layer number 8 for different thicknesses in the additive column. For the $16a$ thickness case, almost perfect crystallization is seen with a negligible fraction of stable dislocation loops. As the thickness of the molten layer is doubled to $32a$, grain boundaries emerge and a polycrystal structure is seen, delineated by grain boundaries (identified as amorphous/white regions). The first grain contains vertical stacking faults while subsequent grains contain slanted twin boundaries, or a combination of stacking faults and twin boundaries.  The larger melt thickness simulated ($48a$) also forms larger grains, that initially contain amorphous regions of high dislocation densities, as seen in the last layer (Fig. \ref{fig:CNA_thickness}(bottom)). These regions would subsequently reform during addition of more layers to form smaller sub--grains. Overall, although different grain sizes are seen for $32a$ and $48a$ cases, the percentage of defects for these two cases are approximately similar.

\begin{figure}[h]
    \centering
    \includegraphics[width=\textwidth]{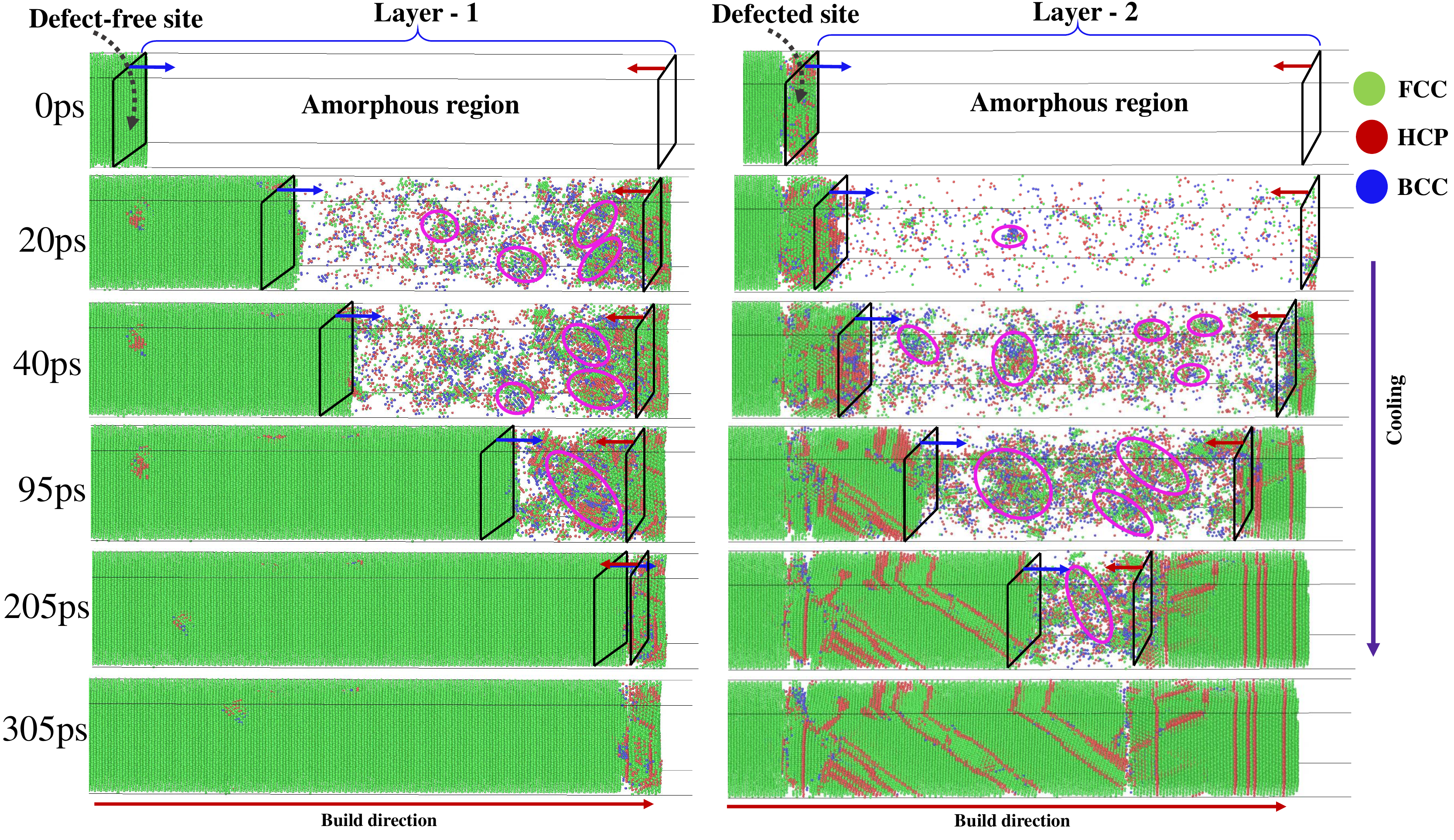}
    \caption{Evolution of two solidification fronts demonstrating the cooling mechanism for a $96a$ thickness case for (left) layer-1 solidification with a perfectly crystalline substrate; and (right) layer-2 which nucleates from the defects left over at the top of layer-1. }
    \label{fig:contourL1L2CNA_thickness}
\end{figure}

The results seen here can be interpreted by studying the solidification kinetics in the melt. The solidification is primarily seen to proceed along two directions for an added molten layer as shown in Fig. \ref{fig:contourL1L2CNA_thickness} for the $96a$ thickness case. The molten layer in contact with the cooler substrate lattice has a fast solidification front that sweeps across the molten layer. The crystal orientation of the substrate is maintained in general. The solidification front at the top, in general, does not have the same crystal orientation as the substrate and thus, when the two fronts interact, grain boundaries are formed. 

\begin{table}[h]
\centering
\caption{The value of solidification front speed at the bottom and top of the melt pool estimated for $96a$ thickness case.}\label{frontspeed} 
\begin{tabular}{P{6.5cm}P{2.5cm}P{2.5cm}P{2.5cm}}
\hline
\multicolumn{1}{c}{\textbf{Solidification front}} & \multicolumn{1}{c}{\textbf{$F^{crystal}_{bottom}$}} & \multicolumn{1}{c}{\textbf{$F^{defect}_{bottom}$}} & \multicolumn{1}{c}{\textbf{$F^{free}_{top}$}} \\ \hline
\textbf{Solidification front speed} [\r{A}/ps] & $1.79$ & $0.768$ & $0.494$ \\ \hline
\end{tabular}
\end{table}

For small layer thicknesses, defect--free single crystals are obtained as the faster solidification front reaches the top of the melt pool before the initiation or within early stages of initiation of the slower front from the top. The critical layer thickness can be analytically found using the speeds of the solidification fronts as follows. The solidification speeds as measured from the case in Fig. \ref{fig:contourL1L2CNA_thickness} is tabulated in Table \ref{frontspeed} (refer animation03 and animation04 in SI for layer-1 and layer-2, respectively). The solidification speed from a defect free substrate is $v = 1.79$ \r{A}/ps. The front speed at the top layer is much lower, at $0.494 $ \r{A}/ps. The crystal structure for subsequent layers begin to form stable crystals only after a delay of $\tau \approx 40$ ps compared to the faster front. To identify the threshold thickness to achieve a single crystal, one needs to ensure that the fast solidification front reaches the top before the crystal (with a different orientation) has a chance to fully initiate at the top. This threshold thickness ($t_{critical}$) can be calculated as $t_{critical} = v\tau$ which is about $20a$, the threshold thickness as seen in Fig. \ref{fig:contour3D_thickness} for the formation of defect free single crystals.  Fig. \ref{fig:CNA_thickness} shows the snapshot of the FCC distribution at the end of cooling of the $8^{th}$ layer for $16a$ thickness case which shows the single crystal structure below the threshold. 

Increasing the thickness of the melt beyond this critical thickness leads to an increase in the number of defects due to interactions of the two solidification fronts. The presence of defects on the substrate further slows down the primary solidification front. As seen in Fig. \ref{fig:contourL1L2CNA_thickness}, if the substrate contains a large number of defect structures (layer 2 as shown at 0 ps), the solidification front velocity moves at a slower pace ($v = 0.768$ \r{A}/ps compared to $v=1.79$ \r{A}/ps for solidification from a defect free substrate) but always faster than the solidification front that moves downward from the top of the melt. This would mean that the solidification fronts meets somewhere closer to mid--way for layer 2 compared to layer 1 as seen in Fig. \ref{fig:contourL1L2CNA_thickness} and forming a grain boundary. The slower speed of the fronts from defective substrate also implies that it takes more time to achieve stable defect structures as the number of layers increase, a feature previously seen in Fig. \ref{fig:fcc_Vs_time16x16x16}.

In addition to the interaction between the two solidification fronts, each front interacts with a homogeneous nuclei (shown as pink ellipsoids) formed within the melt as shown in Fig. \ref{fig:contourL1L2CNA_thickness}. These nuclei when absorbed into the two larger solidification fronts leave defect remnants. For smaller thicknesses (eg. $32a$ case shown in Fig. \ref{fig:CNA_thickness}), these nuclei do not have time to grow as the solidification front rapidly covers the melt. However, the cases with larger thicknesses (eg. $48a$ case shown in Fig. \ref{fig:CNA_thickness}) have significantly large chunks of amorphous regions as the solidification fronts interact with larger and more differentiated internal nuclei that would take longer to coalesce into the primary crystallized region. These regions have higher interfacial energies that are relieved as the additive process proceeds through formation of fine sub--grains.

\begin{figure}[h]
    \centering
    \includegraphics[width=\textwidth]{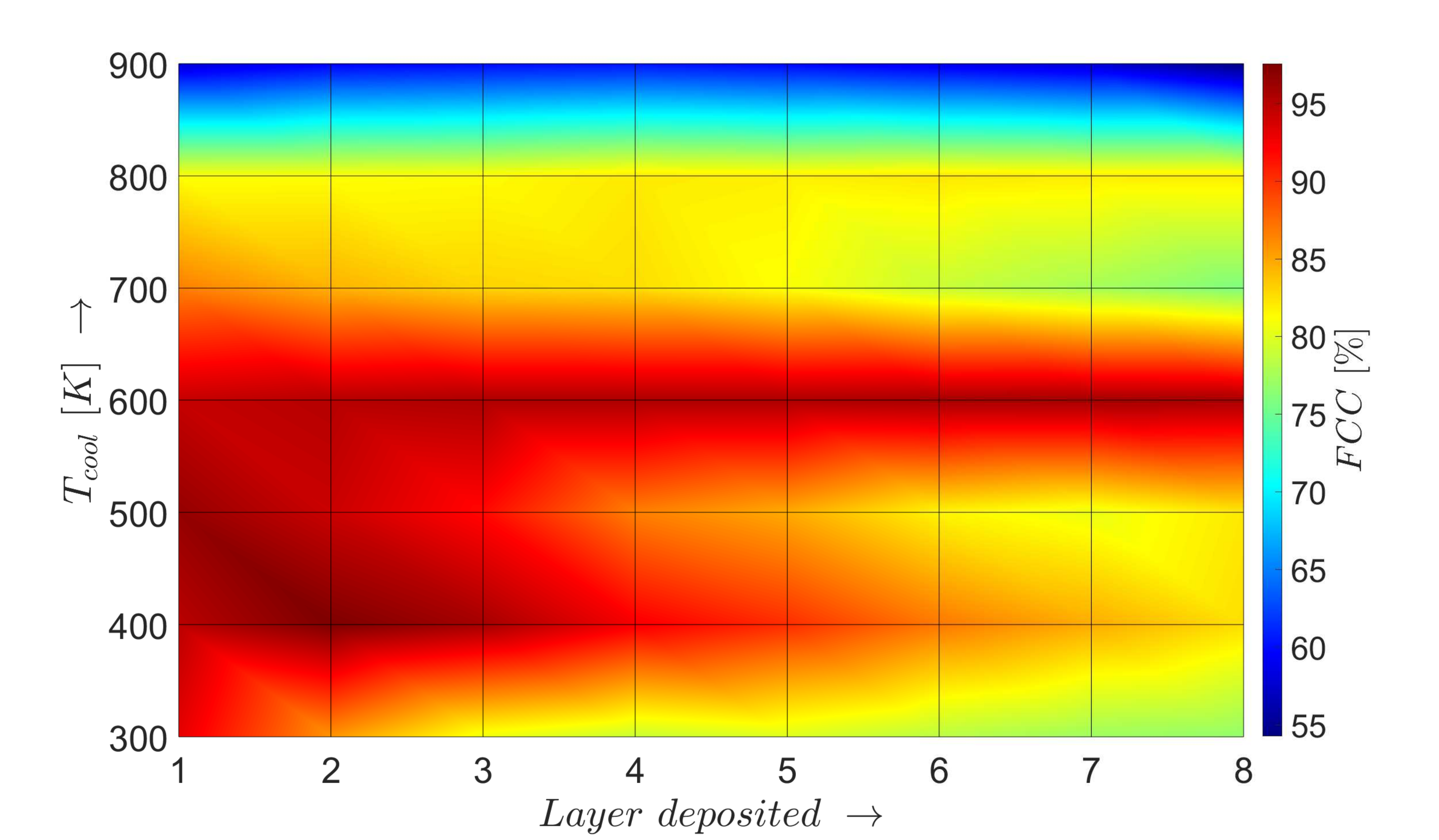}
    \caption{FCC structure percentage obtained as a function of bed temperature and number of layers deposited. A cooling time of 500 ps and a $32a$ layer thickness is used.}
    \label{fig:ContourTcool}
\end{figure}

\subsection{Effect of pre--heated beds (variation of parameter $T_{cool}$)}
In the study outlined in the previous section, each layer was cooled down to room temperature $T_{cool}=300$ K. Experimentally, it is known that preheating the powder beds to higher temperatures leads to less defects in the final product. For example, electron beam melted (EBM) parts has lower microcracks compared to selective laser melted(SLM) parts primarily owing to the lower cooling rate due to preheated powder beds and a vacuum chamber that dissipates heat away slower. In the molecular simulations, preheated beds can be simulated by increasing parameter $T_{cool}$. For this section, we cool down each layer of the additive column to a temperature $T_{cool}$ from 300 K to 900 K with an increment of 100 K. A layer thickness of $32a$ is chosen for the simulation.

\begin{figure}[h]
    \centering
    \includegraphics[width=0.9\textwidth]{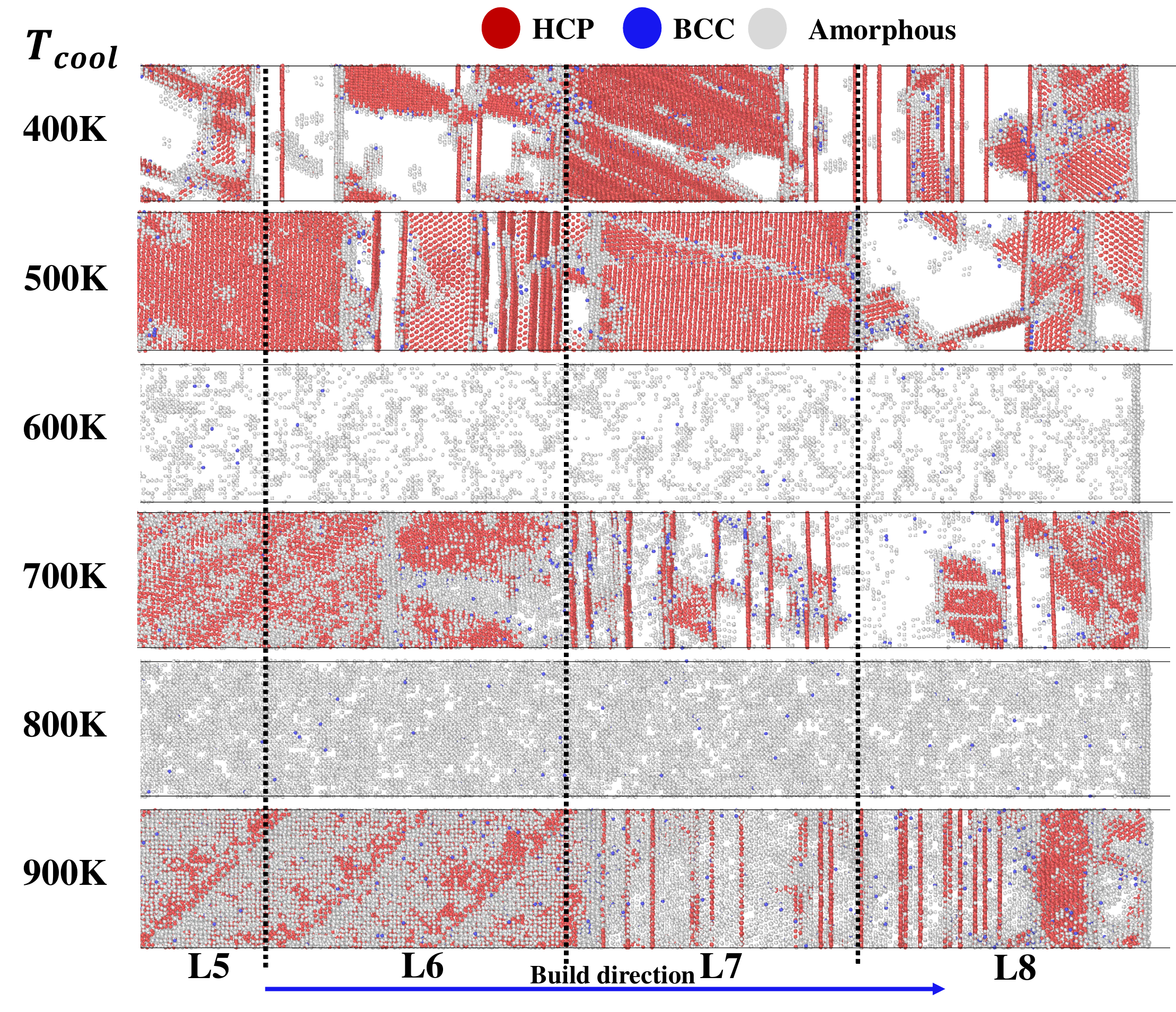}
    \caption{Defect structure after cooling to different substrate temperatures ($T_{cool}$) (only last 4 layers, i.e. $L_5-L_8$ shown).}
    \label{fig:CNADefectsTcool}
\end{figure}

Fig. \ref{fig:ContourTcool} plots percentage of FCC content for different layers as a function of $T_{cool}$. Starting from a higher defect content at $300$ K, the defects initially  improve as a function of temperature reaching almost a defect free single crystal at 600 K. As the temperature is further increased, defects again begin to increase significantly with temperature. At higher bed temperatures, for the same time of cooling, the temperature drop per unit time needed is lower which implies slow cooling rates. Slow cooling rates are beneficial to reduce defects as the atoms have more time to rearrange and reduce defects as seen earlier in Fig. \ref{fig:struc_Vs_L16x16x16}. On the contrary, there is an increase in defects beyond 600 K. This unusual behavior can be explained by looking at the defect structures formed at different cooling temperatures.

\begin{figure}[h]
     \centering
     \begin{subfigure}[b]{0.48\textwidth}
         \centering 
         \includegraphics[trim={1.9cm 0 2.0cm 1.5cm},clip=true ,width=\textwidth]{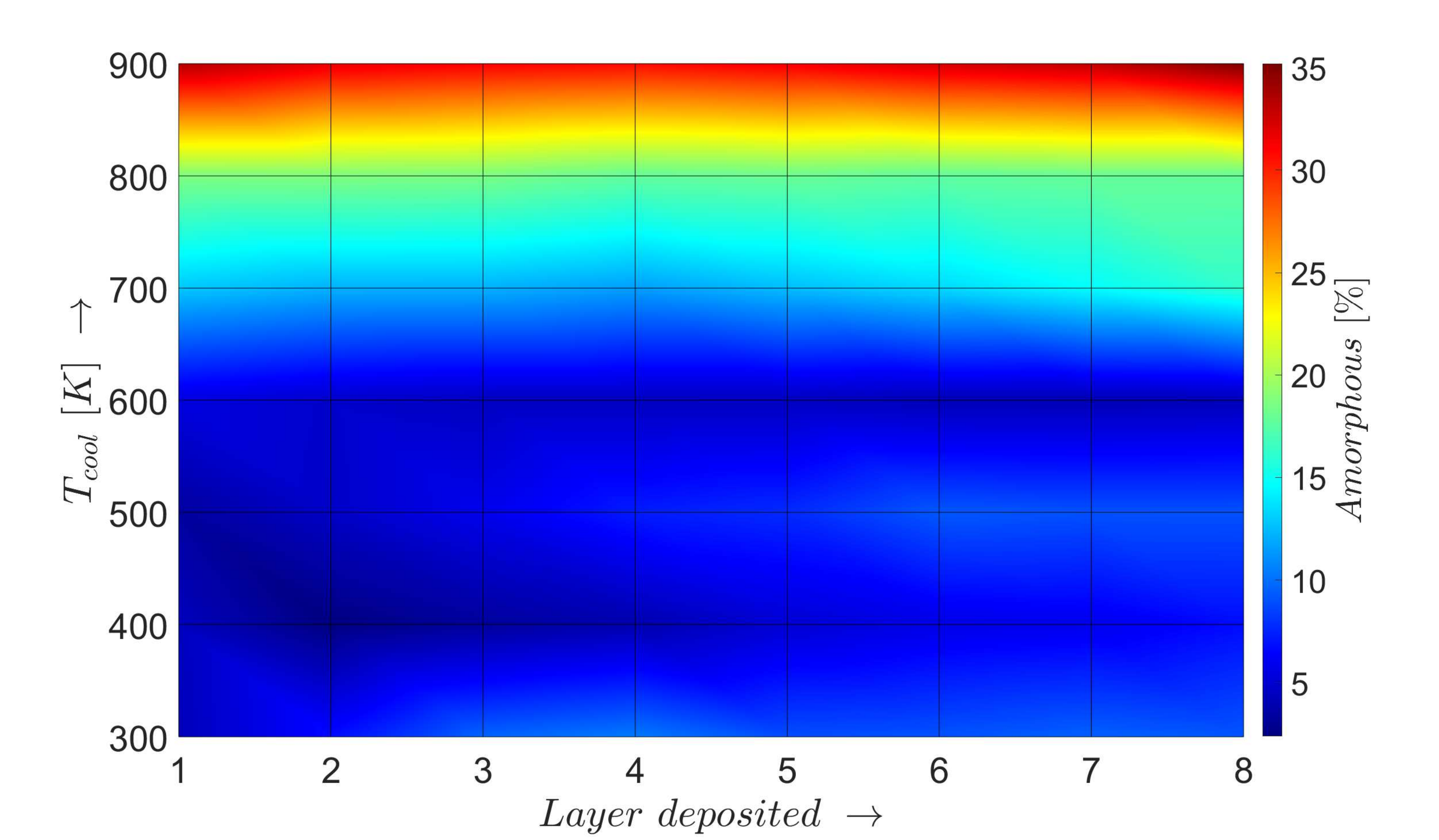}
         \caption{$Amorphous~\%$}
         \label{fig:amorph_Tcool}
     \end{subfigure}
     \hfill
     \begin{subfigure}[b]{0.48\textwidth}
         \centering 
         \includegraphics[trim={1.9cm 0 2.0cm 1.5cm},clip=true ,width=\textwidth]{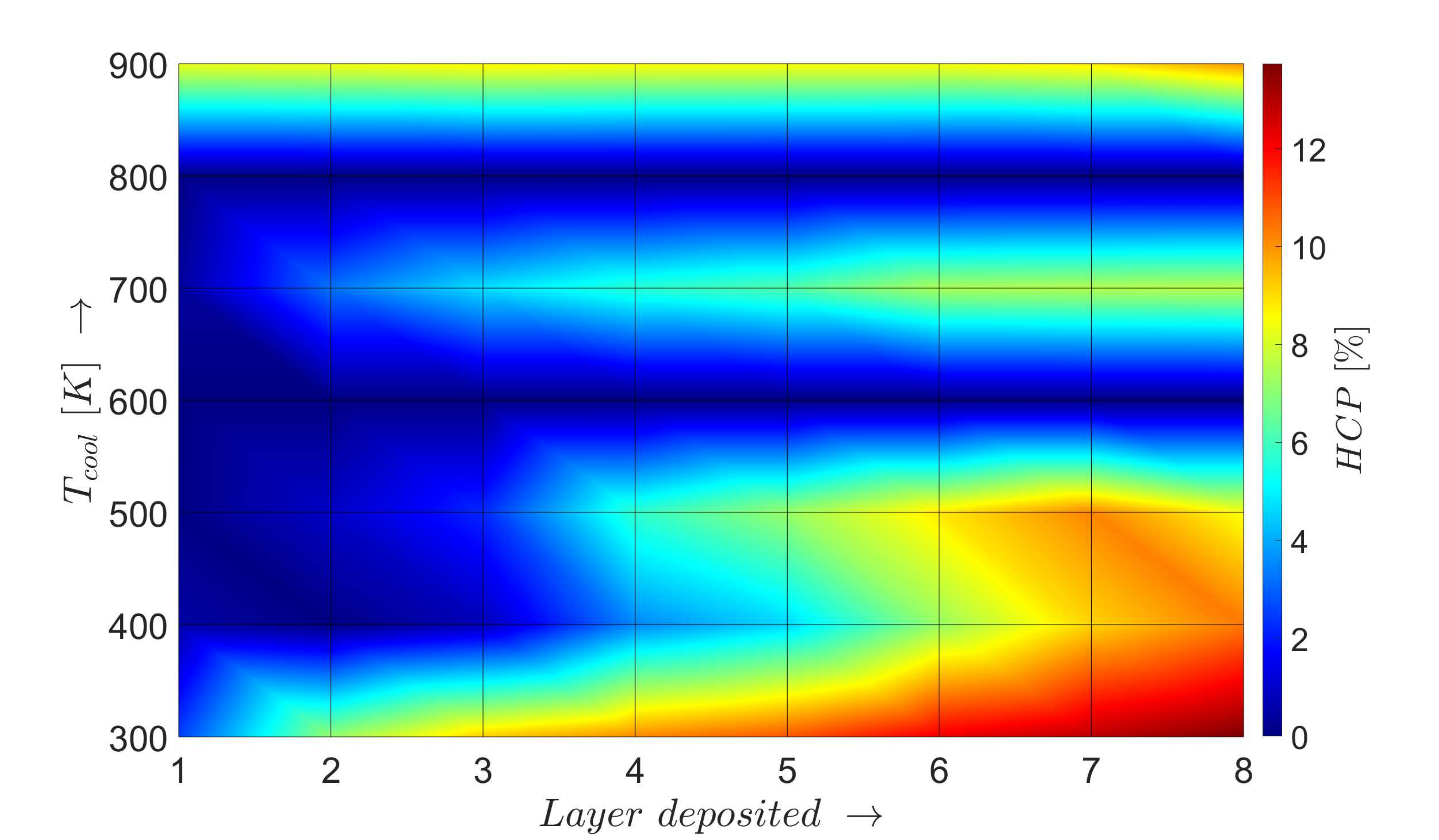}
         \caption{$HCP~\%$}
         \label{fig:hcp_Tcool}
     \end{subfigure}
        \caption{Defect evolution in the additive column under different substrate temperatures with increase in number of layers (a) Percentage of amorphous regions. (b) Percentage of HCP regions.}
        \label{fig:AmorphousHCP_Tcool}
\end{figure}

Fig. \ref{fig:CNADefectsTcool} shows the defect structure for various temperatures of the bed. Only the defects are shown in this figure (and FCC structure is hidden). At 400 and 500 K, most defects are of stacking fault type or grain and twin boundaries. At 600 K all these defects are completely removed, and the defects are of amorphous nature. As the temperature is further increased, the amorphous content begins to significantly increase while a smaller number of stacking faults are seen. Fig. \ref{fig:AmorphousHCP_Tcool} shows the split between amorphous and HCP--type defects as a function of bed temperature for different layers. As seen in Fig. \ref{fig:AmorphousHCP_Tcool}(a), the amorphous content begins to increase from 600 K going toward the melting point of copper. At 900 K, as much as a third of the content is fully amorphous and no benefits of slow cooling rates are seen. Fig. \ref{fig:AmorphousHCP_Tcool}(b) shows that the HCP content increases as the temperature of the bed is decreased below 600 K towards the room temperature. Thus, two mechanisms compete as the bed temperature is raised: (1) slower cooling rates leading to lower stacking fault type defects and (2) high energy amorphous content as one goes closer to the melting point. The interaction between these mechanisms gives a sweet spot for the temperature of the preheated beds for copper at 600 K.

\begin{figure}[h]
    \centering
    \includegraphics[width=0.8\textwidth]{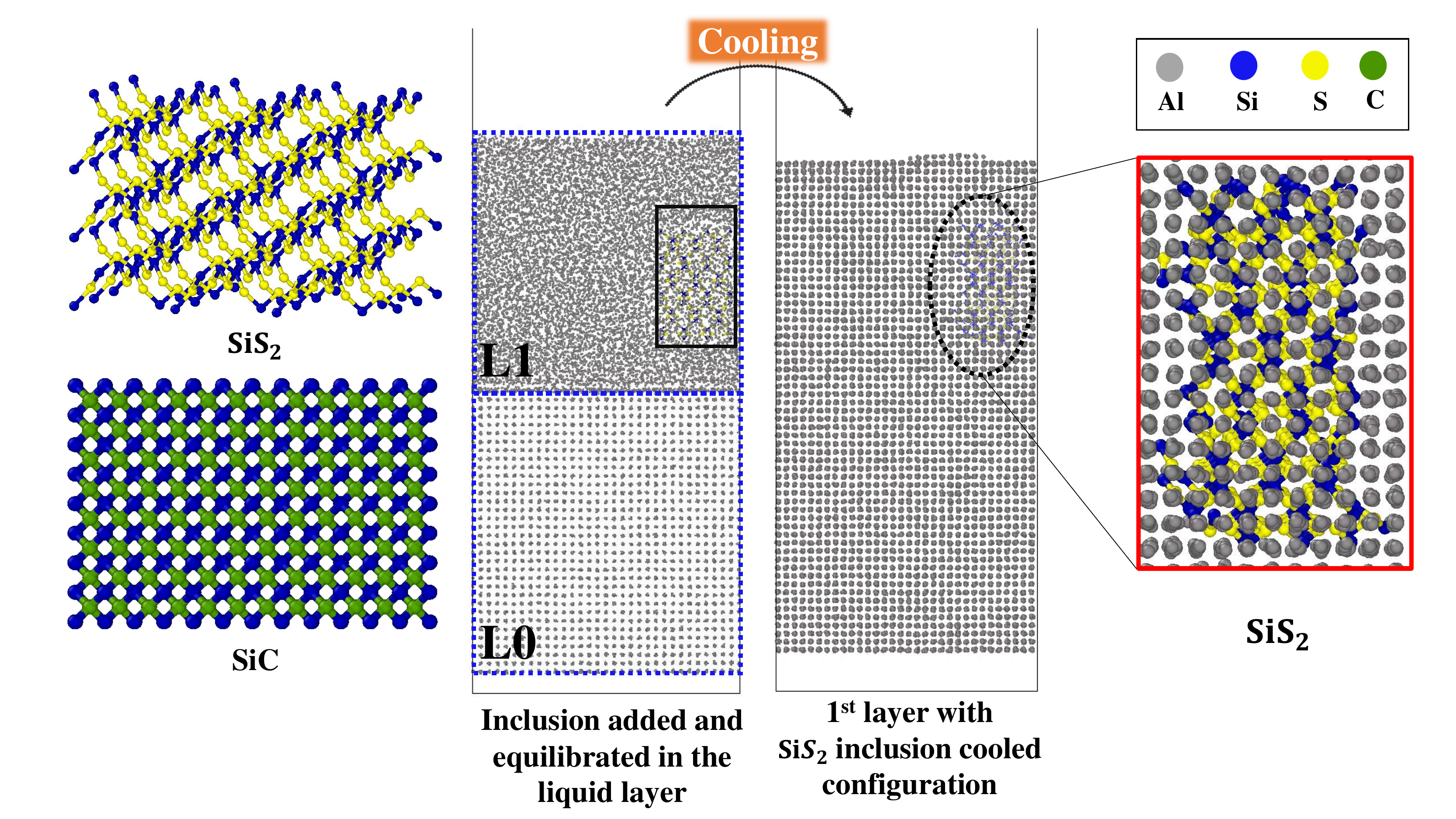}
    \caption{(left) Inclusion SiS$_2$ and SiC (right) SiS$_2$ as solidified in first layer of Al matrix after cooling down ($V_f=2.76\%$ for SiS$_2$ and $V_f=2.88\%$ for SiC)}
    \label{fig:SetupInclusion}
\end{figure}

\subsection{Effect of inclusions}

In this section, the effect of inclusions on atomistic scale defect evolution is studied. It is known that the use of high temperature stable, soft inclusions lead to improved toughness in the final structure, for example, multi-component oxide, oxy-sulphide or sulphide inclusions in iron welds \cite{koseki2005overview}. Two types of inclusions were chosen: a soft inclusion (silicon disulfide, melting point of 1363 K and elastic modulus of 32 GPa) and a hard inclusion (silicon carbide, melting point of 3003 K and elastic modulus of 323 GPa). To simulate inclusions in the melt, the melting point of the inclusions should be higher than the substrate such that the inclusion is retained in the melt. Because of the higher melting point of copper (1358 K), an Aluminum substrate  (melting point of 933 K) was chosen for this example. The structure and density of the constituents are additionally listed in Table \ref{CrystalProp}. Consistent valence force field (CVFF) \cite{dauber1988structure} was used for the bonded interactions of Si$S_2$ and SiC inclusions. Aluminum lattice was modelled using the EAM potential as described in \cite{zhou2004misfit}.

\begin{table}[h]
\centering
\caption{Crystal properties of Al matrix, SiS$_2$ and SiC inclusions }\label{CrystalProp} 
\begin{tabular}{P{2.0cm}P{2.0cm}P{4.0cm}P{3.0cm}P{3.5cm}}
\hline
\textbf{System} & \textbf{Structure} & \textbf{Elastic Modulus} & \textbf{Melting point} & \textbf{Density at RT}\\ \hline
\textbf{Al} \cite{sin2002ab} & FCC & 70.2 GPa  & 933 K & 2.70  gcm$^{-3}$\\ \hline
\textbf{SiS$_2$} \cite{de2015charting} & Tetragonal & 32 GPa & 1363 K  & 2.20  gcm$^{-3}$ \\ \hline
\textbf{SiC} \cite{de2015charting} & Cubic & 323 GPa  & 3003 K & 3.17 gcm$^{-3}$ \\ \hline
\end{tabular}
\end{table}

Leonard-Jones (LJ) potential \cite{dauber1988structure} is used for non-bonded interaction of S-Al interaction for SiS$_{2}$ inclusion simulations.  Al-SiC systems have been studied with Morse potential for interface fracture \cite{Dandekar2011} and interface properties \cite{Luo}. The Morse pairwise interactions energy is computed as follows:
\begin{equation}
    E = D_o\left[e^{-2\alpha(r-r_o)} -2e^{-\alpha(r-r_o)} \right] ~~~~~~ r<r_c
\end{equation}
where, $r$ is the distance between two particles, and the cut-off distance $r_c=10$ \r{A}. The other parameters for the Al-Si and Al-C non--bonded interactions using Morse potential, were originally obtained by Zhao et al. \cite{Zhao2008} and are summarized in Table \ref{ParamMorse}.

\begin{table}[h]
\centering
\caption{Values of the Morse potential parameters obtained by applying inverse method to $ab~initio$ data \cite{Zhao2008}.}\label{ParamMorse} 
\begin{tabular}{P{5cm}P{5cm}P{5cm}}
\hline
\multicolumn{1}{c}{\textbf{Pair}} & \multicolumn{1}{c}{\textbf{Parameter}} & \multicolumn{1}{c}{\textbf{Value}} \\ \hline
\textbf{} & $D_o$ & $0.4824$ eV \\
\textbf{Al-Si} & $\alpha$ & $1.322$ \r{A}$^{-1}$\\
\textbf{} & $r_o$ & $2.92$ \r{A} \\ \hline
\textbf{} & $D_o$ & $0.4691$ eV \\
\textbf{Al-C} & $\alpha$ & $1.738$ \r{A}$^{-1}$ \\
\multicolumn{1}{l}{} & $r_o$ & 2.246 \r{A}\\ \hline
\end{tabular}
\end{table}

First the equilibrated density of pure Al at 1100 K is obtained by running NPT simulation at 1 bar pressure conditions and the density was observed to stabilize to $\rho_{liquid}=2.297$ gcm$^{-3}$. Each layer of this melt density at 1100 K was added to the additive column. In order to insert the inclusion, atoms in the amorphous layer (with a volume equivalent to the volume and shape of the inclusion) are carved out by deletion from this melt pool layer. Then, the melt with the inclusion is again equilibrated at 1100 K at which point Aluminum melts while the inclusion is retained. The same process as explained in the {'}methods{'} section is repeated for modeling sequential addition of the melt. The system is cooled to room temperature with a cooling time of 100 ps.  An Al $16a\times16a\times16a$ thickness system is chosen for this study and a single inclusion is added at each layer. Fig \ref{fig:SetupInclusion} depicts the inclusions SiS$_2$ and SiC in the first layer of solidified Al melt after cooling down. Volume fractions of the inclusion for the two cases studied are similar with $V_f=2.76\%$ for SiS$_2$ and $V_f=2.88\%$ for SiC inclusion. 

\begin{figure}[h]
     \centering
     \begin{subfigure}[b]{0.4\textwidth}
         \centering
         \includegraphics[width=\textwidth]{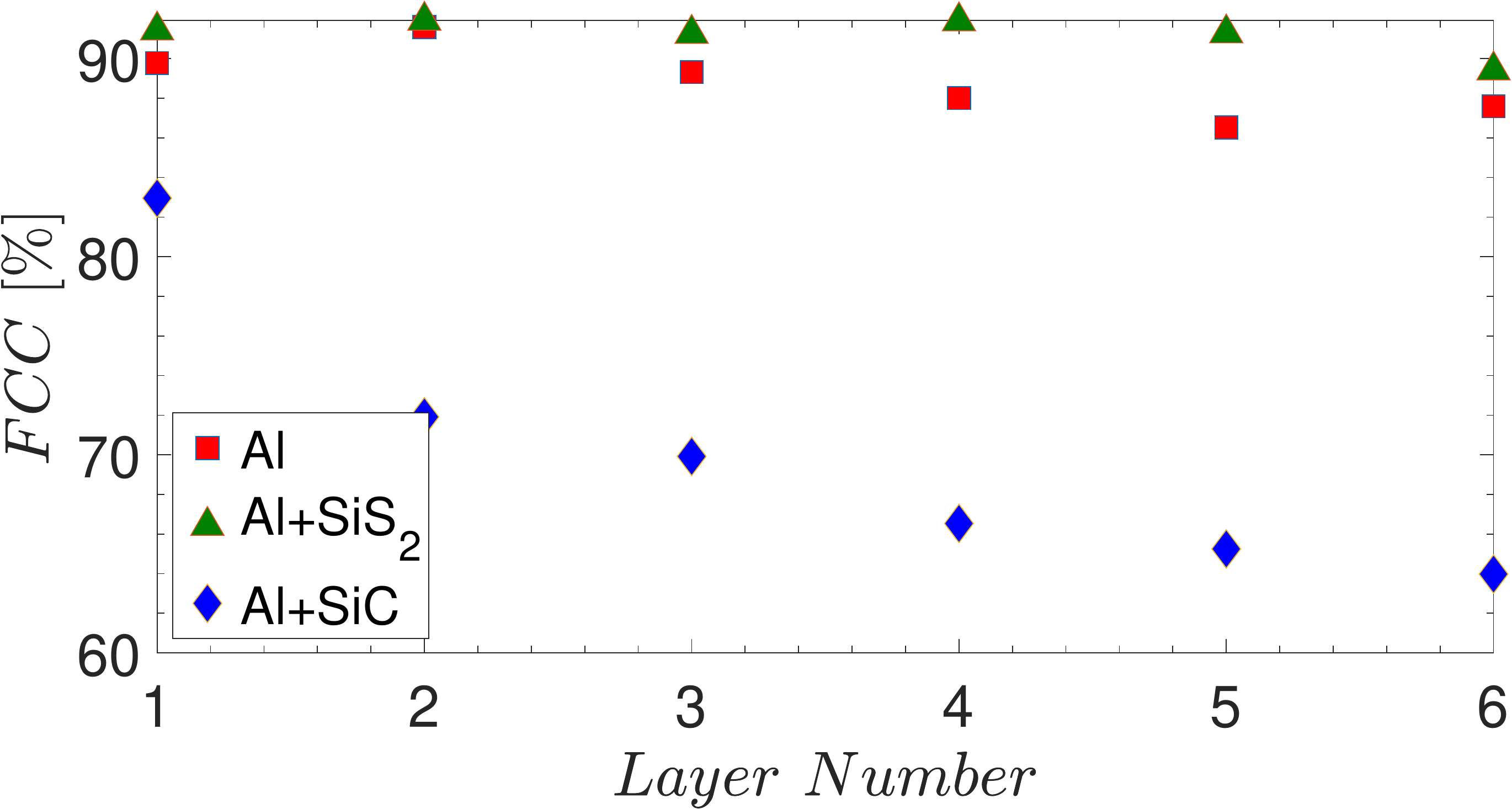}
         \caption{$FCC~\%$}
         \label{fig:FCC_Al_inclusion}
     \end{subfigure}
     \hfill
     \begin{subfigure}[b]{0.5\textwidth}
         \centering
         \includegraphics[width=\textwidth]{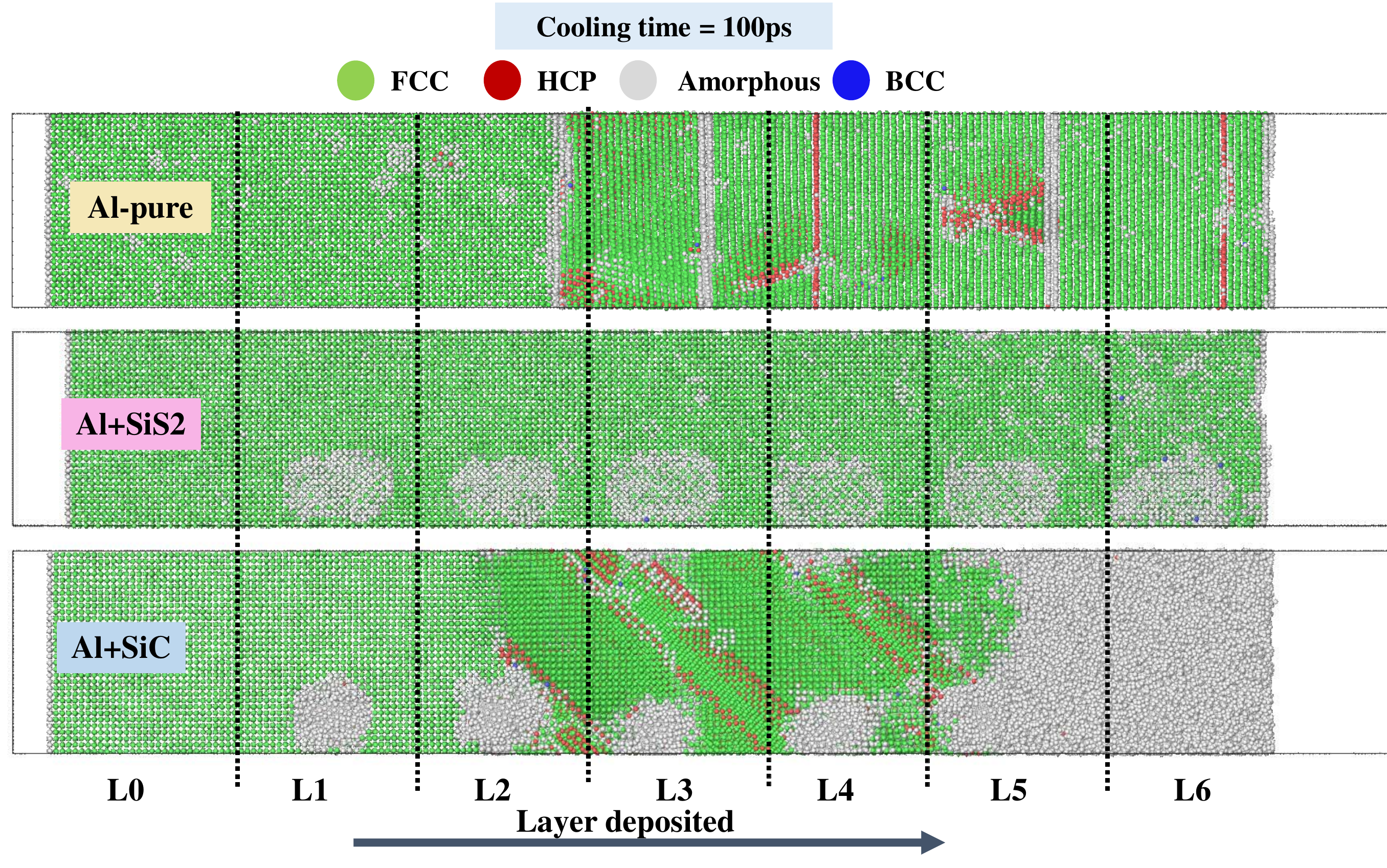}
         \caption{CNA snapshots}
         \label{fig:CNA_Al_inclusion}
     \end{subfigure}
        \caption{(a) FCC percentage evolution as a function of deposited number of layers for pure Al compared against Al with two inclusions studied here and (b) defect snapshots after deposition of six layers for $16a$ layer thickness.}
        \label{fig:Inclusion}
\end{figure}

Fig. \ref{fig:FCC_Al_inclusion} shows the influence of the soft (SiS$_2$) and the hard (SiC) inclusion on defect evolution as compared to a process with pure Al. The soft inclusion showed the lowest defect percentage and the hard inclusion showed significant amount of defects. Fig. \ref{fig:CNA_Al_inclusion} shows the defect structure after the sixth melt layer is cooled.  Pure Al forms a polycrystalline structure with grain boundaries (amorphous vertical regions) under these conditions with a few retained stacking faults. With the softer inclusion, a single crystal orientation is retained and stacking faults are eliminated.

Solidification velocities of the pure Al and Al with SiS$_2$ are found to be similar, showing that the  SiS$_2$ inclusion does not inhibit or enhance solidification front speeds. However, it partially shields interaction between the competing solidification fronts.  On the other hand, solidification fronts move much slower in the case of a harder SiC inclusion.  In this case, large amorphous regions are formed around the precipitate. This amorphous region is seen at layers 5 and 6 in Fig. \ref{fig:CNA_Al_inclusion} (bottom), but also formed when layers 1 to 4 were added. These regions subsequently reformed to a twinned region emanating from the precipitate (as seen in layers 2-4). Due to the lack of defects, use of softer inclusions in metal powders is worth pursuing in the future to achieve products with improved fracture properties.

\section{Conclusions}
In this work, we use molecular dynamics simulations to systematically model a nanoscale additive column to investigate the influence of layer thickness, cooling time, target cooling temperature and alloy inclusions on the final defect structure. Such simulations, although idealizing a very complex additive manufacturing process, are able to capture the non--equilibrium physics at nanoscales that lead to formation of defect structures. It is found that the percentage of defect--free content of copper converges when a sufficient number of layers are added. The solidification is primarily seen to proceed along two directions for an added molten layer. The molten layer in contact with the cooler lattice has a fast solidification front that competes with the slower solidification front starting from the top layer. Defect structure formed strongly depends on the interactions between these competing solidification fronts. The key observations from these simulations are as follows:

\begin{itemize}
    \item Slower cooling rates lead to a reduction in defects, however, the benefits diminish below a critical rate as stable dislocation defects form that cannot be further eliminated.
    \item Up to a critical layer thickness, defect free single crystals are obtained as the faster solidification front reaches the top of the melt pool before the stable formation of the slower front from the top. 
    \item As the thickness of the molten layer increases beyond a critical thickness, grain boundaries emerge and a polycrystalline structure is formed. The grain sizes typically increase with the layer thickness.  However, for large melt thicknesses, amorphous regions of high dislocation densities are formed as the solidification fronts interact with more differentiated homogeneous nuclei. These regions subsequently reform to form smaller sub--grains.
    \item Defect content can be significantly reduced by raising the temperature of the powder bed to a critical temperature. At low bed temperatures, faster cooling rates lead to significant dislocation defects. At higher temperatures, the cooling rates are lower leading to lower dislocation content. However, higher temperatures also lead to an increase in  amorphous content. There is a critical value of temperature that balances both the dislocation defects and the amorphous content. 
    \item Finally, the effect of added soft inclusion (SiS$_2$) and a hard inclusion (SiC) on the defect structure in Aluminum is studied. SiC inclusion significantly slows down the solidification front leading to retained defect structure. However, addition of SiS$_2$ does not modify the solidification velocity compared to pure Al. Additionally, presence of SiS$_2$ is seen to reduce defective content compared to a pure metal. 
\end{itemize}

The presented results are an initial step towards computational understanding of the additive process parameter--crystal structure relationship in a non--equilibrium setting. This study can be improved by considering the effect of actual laser heating profiles, surface cooling in a non vacuum environment and interaction of melt pools. The methodology can also be used to model evolution of residual stresses in the unit cell as a function of process parameters as well as simulate the stress--strain response after processing. A critical advantage of a first principles approach is that the simulations can be used to perform alloy design, as the preliminary inclusion case presented here, but with improved modeling of formation and dispersion of multiple inclusions.

\section*{Acknowledgement}
This research was supported in part through computational resources and services provided by Advanced Research Computing at the University of Michigan, Ann Arbor. The authors would like to thank Mr. Aaditya Lakshmanan for his valuable discussions.

\section*{Supplementary Information}
Supplementary material contains animations of the simulations of the additive column crystallization and defect evolution during cooling of different layers under various conditions.

\noindent \textit{Data Availability Statement:} The data that support the findings of this study are available from the corresponding author upon reasonable request.

 \section*{References}

\bibliography{reference_list}
\end{document}